\documentclass[aps,prd,notitlepage,preprintnumbers,superscriptaddress,nofootinbib,11pt]{revtex4-2}
\usepackage[utf8]{inputenc}
\usepackage[a4paper, total={7in, 9in}]{geometry}
\usepackage{amsmath}
\usepackage{amssymb}
\usepackage{enumerate}
\usepackage{amsmath}
\usepackage{tikz}
\usepackage[compat=1.1.0]{tikz-feynman}
\usepackage{feynmf}
\usepackage{slashed}
\usepackage{braket}
\usepackage{url}
\usepackage{natbib}
\usepackage{graphicx}
\usepackage[pdftex,bookmarks,linktocpage,pdfpagelabels,plainpages=false,hyperfigures,linkcolor=blue,citecolor=blue,urlcolor=blue]{hyperref} 
\hypersetup{colorlinks=true}

\usepackage{mathrsfs,amssymb}  
\usepackage{cancel}
\usepackage[normalem]{ulem}

\usepackage{orcidlink}

\usepackage{makecell}
\usepackage{tabularx}
\newcolumntype{Y}{>{\centering\arraybackslash}X}

\usepackage{fontawesome}
\usepackage{xcolor}
\usepackage{xspace}
\usepackage{makecell}

\begin{document}

\title{The {\underline {DA}}rk {\underline M}essenger {\underline S}earches at an {\underline A}ccelerator Experiment, \\
A Case of a Table-Top Scale Experiment at a Beam Dump}
\date{31 March, 2025}
\collaboration{DAMSA Proto-collaboration}
\author {Jaehoon Yu} 
\email{jaehoonyu@uta.edu}
\affiliation{Dept. of Physics, University of Texas, Arlington, TX76019, USA}
\author {Doojin Kim}
\email{doojin.kim@usd.edu}
\affiliation{Dept. of Physics, University of South Dakota, Vermillion, SD 57069, USA}

\author {Un-Ki Yang}
\email{ukyang@snu.ac.kr}
\affiliation{Dept. of Physics and Astronomy, Seoul National University, Seoul, 08826, Republic of Korea}

\maketitle
\clearpage

\begin{center}
{\bf {\underline{\underline{EXECUTIVE SUMMARY}}}}
\end{center}
 DAMSA ({\bf{\underline {DA}}}rk {\bf{\underline {M}}}essenger {\bf{\underline {S}}}earches at an {\bf{\underline {A}}}ccelerator) is a table-top scale, extremely-short-baseline experiment designed to probe dark-sector particles (DSPs) that serve as portals between the visible sector and the hidden dark-matter sector. These particles, such as axion-like particles (ALPs), can decay into two photons or $e^{+}e^{-}$ pairs. DAMSA is specifically optimized to explore regions of parameter space that are inaccessible to past and current experiments, by operating at ultra-short baselines and employing high-resolution calorimetry, precision timing, and precision tracking in a magnetic field with suppression of beam-related neutron backgrounds.
The experiment can be integrated into facilities such as CERN’s Beam-Dump Facility (BDF), operating concurrently with the SHiP experiment, and provides complementary sensitivity in the MeV to GeV mass range. DAMSA represents a cost-effective and timely opportunity to expand CERN’s discovery potential in dark-sector physics. It exemplifies how innovative, small-scale experiments can effectively complement large-scale experiments, taking advantage of existing and future infrastructure.
\section{Introduction}

\label{sec:Introduction}
The precision measurements of neutrino properties needed to modify the Standard Model (SM) to reflect the non-zero neutrino mass require high flux beams of neutrinos and large active mass detectors. Future long-baseline neutrino experiments, such as the Deep Underground Neutrino Experiment (DUNE)~\cite{DUNE:2020lwj} in the U.S. and Hyper-K~\cite{hk-cdr} in Japan, are under construction for this purpose. These experiments utilize high-power proton accelerators to produce high-intensity neutrino beams and have 40~kt liquid argon or 180~kt pure water active mass far detectors, respectively. 

Fermilab’s PIP-II LINAC~\cite{PIP2_CDR} currently under construction is a crucial component to DUNE and designed to deliver up to 2~mA total current of 800~MeV protons, of which 1$\sim$2\% would be used to support 1.2~MW beam power to DUNE and other programs at the lab. 
The 2023 U.S. Particle Physics Project Priority Panel report~\cite{p5-2023} based on the U.S. Snowmass HEP strategy studies acknowledges and emphasizes taking advantage of the excess beam capacity of the PIP-II LINAC.  Several subsequent workshops held at Fermilab and multiple discussions with the leadership resulted in the commissioning of the task force for a beam dump facility, tentatively named as Fermilab Facility for Dark sector Discovery (F2D2).  The task force recently submitted a report to the Fermilab leadership.  DAMSA is one of the proposed experiments, which could be placed in front of the large shielding block with its own small target for a short baseline.

CERN, on the other hand, made a decision to construct BDF at ECN3, taking advantage of the high flux 400~GeV SPS proton beams.
The SHiP experiment~\cite{ship-facility} will be sited in the facility which has some space that could also host other experiments, such as SND, in front of it.
The current design of CERN BDF has sizeable shielding materials of the order 20~m, including the magnetic muon shield, after the beam dump.
While the space in front of the SHiP experiment is insufficient for a large-scale experiment, a small desk-top scale experiment, such as DAMSA could be placed in an optimal position immediately behind the muon magnetic shield.

This white paper describes the proposed DAMSA experiment which has been developing over the past three years and has formed a proto-collaboration to pursue exploring the DSP parameter space inaccessible thus far.  While the extreme short-baseline nature of the experiment will surely pose challenges, especially in the harsh radiation environment in the high flux proton beam accelerators, we describe possible ways to overcome the challenges and the theoretical bases for the reason why we are convinced that DAMSA can be successful.

\section{Dark Sector Particle Physics Motivation}
A compelling motivation for new physics comes from the existence of dark matter comprising about 25\% of the university's energy budget, which is strongly supported by numerous astrophysical and cosmological observations through its gravitational effects. However, its particle properties, including the mass scale, remain unknown. Scenarios involving weak-scale weakly interacting massive particles (WIMPs) have garnered significant attention as these candidates are thermally produced, independent of specific cosmological model details before the relic density is determined, and naturally exist in a variety of well-motivated new physics theories beyond the Standard Model.
Moreover, since WIMPs predict non-gravitational interactions between dark matter and SM particles, extensive experimental and theoretical efforts have been devoted to their study over the past few decades (see, e.g., Ref.~\cite{Bertone:2004pz}). However, no conclusive evidence has been found, prompting the exploration of alternative ideas that are now being actively investigated.

Among these alternatives, MeV-scale (light) dark matter is a promising candidate, as it can be thermally produced and its parameter space remains largely unexplored. Additionally, portal scenarios suggest that other dark-sector particles--such as messenger particles mediating the interactions between dark matter and SM particles--of similar mass should exist, with feeble interactions with SM particles (see, e.g., Refs.~\cite{Holdom:1985ag,Patt:2006fw,Huh:2007zw,Pospelov:2007mp,Pospelov:2008jd,Pospelov:2008zw,Falkowski:2009yz,Chun:2010ve,Arcadi:2019lka,Batell:2022xau}). 
The predicted mass scale is within reach of existing beam facilities, which can produce these dark-sector particles, including MeV-scale dark matter, while the very weak coupling strengths motivate the need for experiments with high-intensity beams to accommodate the rare production of such particles. 
In addition to scenarios involving light dark matter, visibly-decaying messenger particles are often proposed to explain various experimental anomalies, such as the MiniBooNE low-energy excess~\cite{MiniBooNE:2008yuf,MiniBooNE:2018esg,MiniBooNE:2020pnu}, the muon $g-2$~\cite{Muong-2:2006rrc,Muong-2:2021ojo}, and the LSND anomaly~\cite{LSND:2001aii}.

With their powerful beams, the PIP-II LINAC and CERN BDF are expected to produce the aforementioned feebly-interacting messenger particles in abundance, shedding light on dark-sector physics. 
The discovery of dark-sector particles at an accelerator would pave the way for a deeper understanding of the nature of dark matter. In particular, high-intensity beam experiments have provided leading constraints on messenger particles in the MeV-to-sub-GeV range. Schematically, messenger particles produced at the target or dump must survive long enough to reach the detector and decay within its fiducial volume. The sensitivity of an experiment depends on factors such as baseline distance, effective decay volume, beam intensity, and beam energy. A key challenge is probing messenger particles that decay relatively early, as longer baselines reduce detection sensitivity, which is referred to as beam-dump ``ceiling''~\cite{Dutta:2023abe,Kim:2024vxg}. To address this, experiments with very short baselines are better suited for exploring these regions of parameter space~\cite{Dutta:2023abe,Kim:2024vxg}.

This study provides the foundation for the proposed DAMSA experiment, a novel experiment for searching for dark-sector particles, inspired by the experimental scheme proposed in Ref.~\cite{Jang:2022tsp}. An essential element of DAMSA is the beam-dump facility, which can produce dark-sector particles using high-intensity proton beams, such as that from the PIP-II LINAC or at CERN BDF. Given the potential of DAMSA to explore the dark sector, it is crucial to place this effort within the broader scope of dark-matter research. 

\section{DAMSA Baseline Studies and Sensitivity at F2D2}

As described above, DAMSA~\cite{Jang:2022tsp} is a very short baseline beam-dump experiment aimed to probe the parameter space inaccessible in previous beam-dump experiments. 
Given the proximity to the beam dump/target, the primary background comes from the large number of beam-related neutrons (BRN), resulting from the beam interactions in the dump, in particular for the incident proton beams. 
In order to overcome these backgrounds, DAMSA aims to detect two-photon or $e^{+}e^{-}$ final states of dark-sector particles and the benchmark physics case for these final states are the ALPs.

\begin{figure}[t]
    \centering
\includegraphics[width=0.75\textwidth]{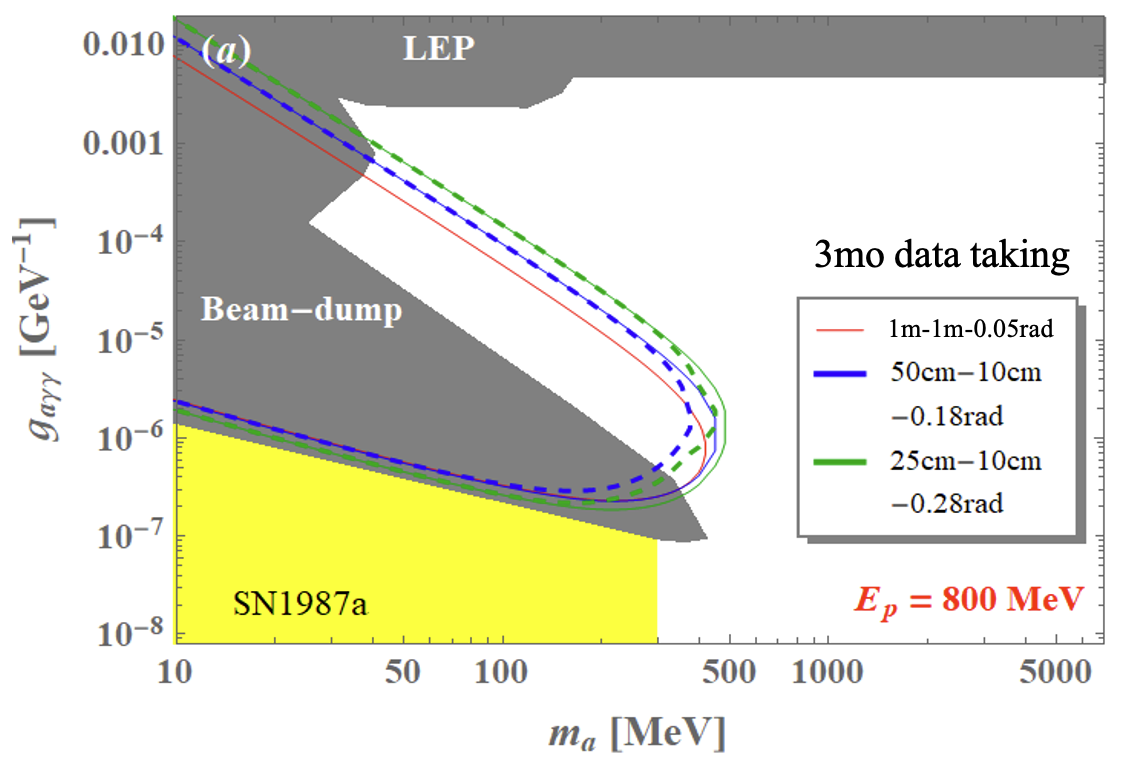}\hspace{2pt}
    \caption{Table-top scale DAMSA sensitivity on the Axion-Like Particles decaying to two photons at Fermilab's PIP-II LINAC's 800~MeV proton beams.  The three lines demonstrates the sensitivity reach for three different distances between the ALP production point and the detector expands the sensitivity further into the unexplored, open parameter space.}
    \label{fig:damsa-sensitivity}
\end{figure}

The initial studies in Ref.~~\cite{Jang:2022tsp} of the background from the anticipated high flux neutrons were performed based on detailed GEANT simulations. Based on this study, the collaboration determined that in order to mitigate the backgrounds from accidental overlaps of the photons from BRN spallation interactions by 10 orders of magnitude, the following detector capabilities are essential:
\begin{itemize}
\item Fine granular calorimeter with high precision shower position resolution to identify the decay vertex better than 1~cm 
\item Excellent timing resolution of the detector for the two EM particle arrival time difference better than sub-ns level
\item As fine a mass resolution as possible, better than an MeV level
\item  Precision tracking system under a magnetic field to distinguish charged particles from photons, identify the sign of the charged particle and measure the momenta
\end{itemize}
The sensitivity of the DAMSA experiment at Fermilab's PIP-II LINAC with the above assumptions on the background and the experiment configuration that varies at three different baseline lengths and the detector angular coverage is shown in Fig.~\ref{fig:damsa-sensitivity}.
The sensitivity reach shows the clear dependence to the distance to the beam source which supports the experimental concept for extremely short baseline, enabling access to the previous unexplored parameter space.
This fundamental tenet of DAMSA is applicable to other physics signatures that result in two clearly identifiable final state particles, assisted by optimal beam configurations and the detector capabilities.

\section{Conceptual Design of a Small Scale DAMSA and Cost Estimate}

\begin{figure}[th]
    \centering
    \includegraphics[width=\textwidth,height=0.3\textwidth]{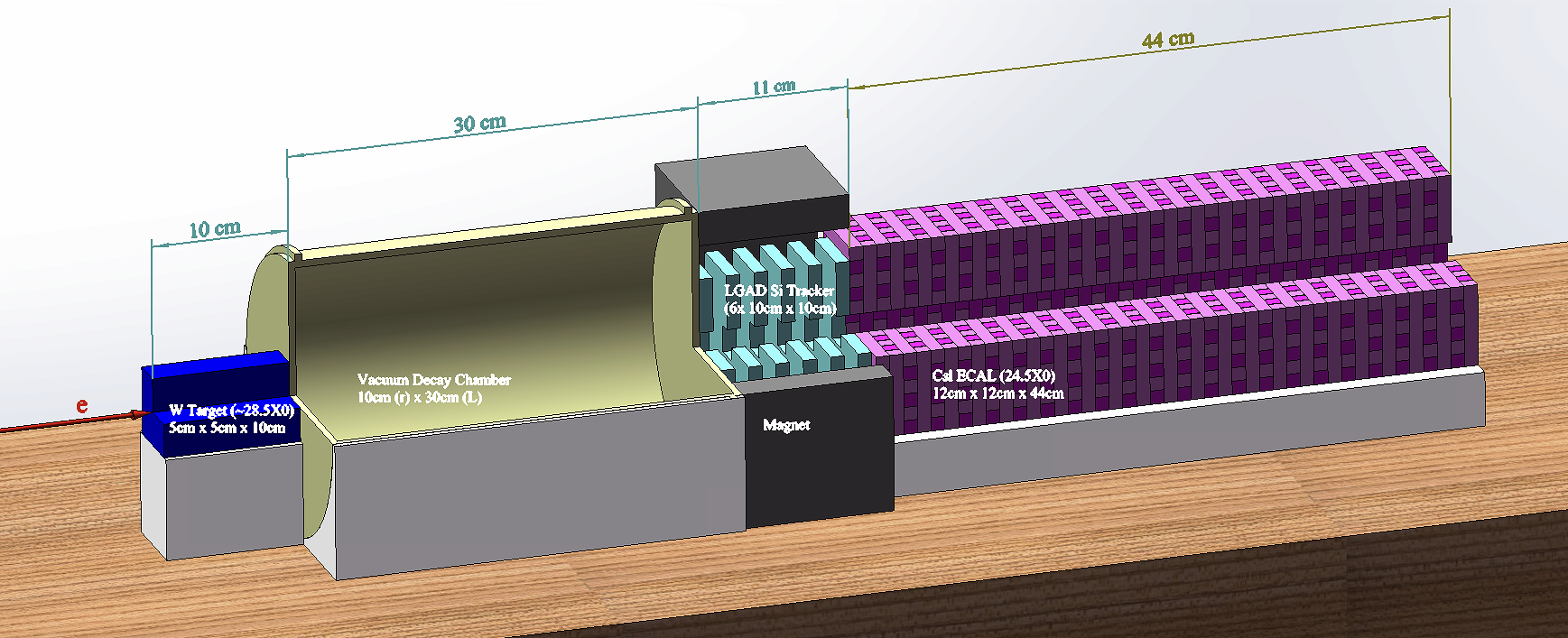}
    \caption{A 3D model of a small scale DAMSA detector. Given the short $10~cm$ target length, a $2~{\rm cm}\times 2~{\rm cm}$ square hole through the entire length of the detector is implemented to protect the detector from the large number of primary protons passing through without interacting in the target. }
    \label{fig:little-damsa-det}
\end{figure}

Based on the above detector requirements and the subsequent detailed studies on the limitations in Ref.~\cite{damsa:ceiling}, the DAMSA collaboration is planning on constructing a compact detector for PIP-II beams as shown in Fig.~\ref{fig:little-damsa-det}.
It has a 2 cm $\times$ 2 cm square hole through the entire length of the detector to minimize the impact of the primary beam protons that exit the target without an interaction.

The experiment consists of a tungsten target of dimension 5 cm (H) $\times$ 5cm (W) $\times$ 10cm (L, 24.5$X_{0}$), a vacuum decay chamber of dimension 10 cm (r) $\times$ 30 cm (L), six layers of LGAD Si tracker under 1T magnet field, provided by a permanent magnet which is employed to minimize infrastructure necessary to support an electromagnet, which would be exposed to a harsh radiation environment.
The six LGAD Si tracker layers have 10cm $\times$ 10cm transverse dimension, each, and two neighboring layers are spaced by a 2~cm center-to-center distance. The total number of readout channels for the LGAD Si tracker system is about 300.

The last detector component is a 4D undoped CsI crystal total absorption EM calorimeter with the dimension 12cm (H) $\times$ 12cm (W) $\times$ 44 cm (L), providing the depth of total $28.5 X_{0}$.
The total absorption 4D electromagnetic calorimeter will be constructed out of layers of small Cesium Iodide crystal bars of dimension 1cm~(W) $\times$ 1cm~(H) $\times$ 12cm~(L) placed in alternating orthogonal layers.  
The total number of readout channels for the CsI ECal in this particular configuration is about 1,000.
This configuration is designed to allow for an accurate reconstruction of the individual photon shower position, energy, and the time of arrival, as well as the projected position of the vertex from which the two decay products originate.
At the time of writing this white paper, however, the collaboration is working on optimizing the geometry of the ECAL, in particular the granularity, to ensure the detector occupancy at the manageable level.
We expect the total cost of the detector is to be less than $\sim$\$10M.

At the time of writing this white paper, the collaboration has worked with the Ukrainian company, the Institute for Scintillating Materials (ISMa) and successfully produced 50 crystals bars of dimension 1cm~(W) $\times$ 1cm~(H) $\times$ 12cm~(L), cut out of one 5cm~(W) $\times$ 5cm~(H) $\times$ 50cm~(L) undoped CsI block previously used in the KTeV experiment at Fermilab.  
These crystal bars have been polished, wrapped with reflective foils and initially quality controlled by ISMa.
They have been distributed to collaborators for further testing.
Both the dimensions of the crystal bars and, therefore, the number of readout channels will naturally evolve, depending on the outcome of the ongoing detailed geometry studies to ensure optimal signal efficiency and background rejection.
We anticipate a conceptual design report of DAMSA by the summer 2025.

\section{DAMSA Sensitivity at CERN BDF}

\begin{figure}[t]
    \centering
    \includegraphics[width=0.5\linewidth]{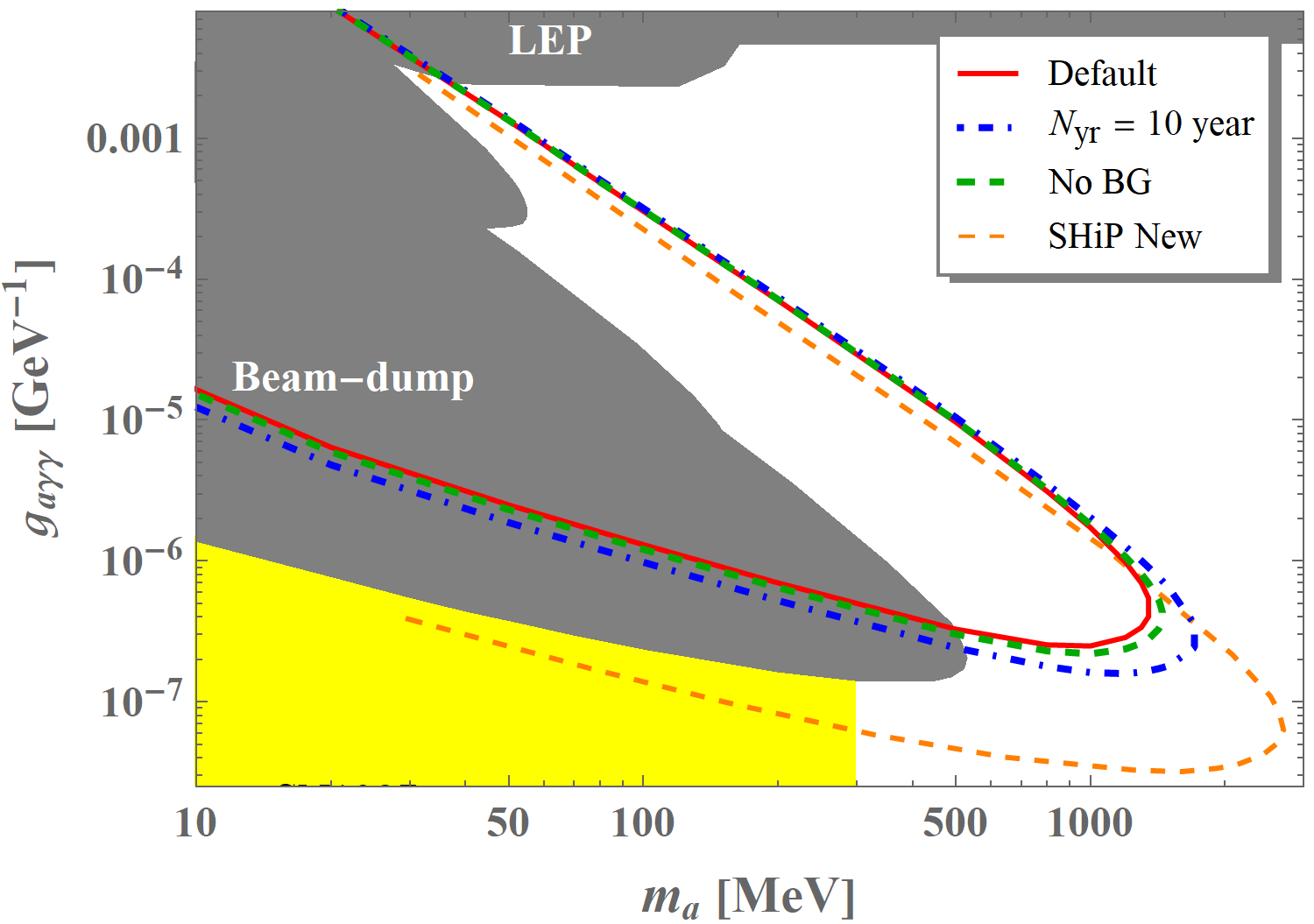}
    \caption{Preliminary sensitivity estimates of DAMSA at the BDF facility.}
    \label{fig:sensitivity}
\end{figure}

For the possibility of DAMSA at CERN's BDF, one can consider excluding the $10~{\rm cm}$ tungsten target from the experiment in Fig.~\ref{fig:little-damsa-det}, since there is going to be a sizable $20~{\rm m}$ dump, equipped with a magnetized muon shield.
Our preliminary sensitivity estimates for DAMSA at CERN BDF, targeting ALP signals interacting with photons, are presented in FIG.\ref{fig:sensitivity} under default assumptions (red solid line). These assumptions include positioning the detector 20 meters from the beam target system, using the $12~{\rm cm}\times 12~{\rm cm}\times 100~{\rm cm}$ electromagnetic calorimeter in Fig.~\ref{fig:little-damsa-det}, a one-year exposure to a $4\times 10^{19}$ POT/year beam, a 5 GeV energy cut on final-state photons, and the background level estimated in Ref.~\cite{damsa:ceiling}. Additionally, we provide sensitivity estimates for a longer exposure (10 years, blue dot-dashed line) and under a background-free assumption (green dashed line). For comparison, the expected sensitivity reach at SHiP~\cite{Albanese:2878604} is shown by the orange dashed line, which clearly demonstrates the large-scale angular coverage of the experiment, extending the covered mass range and the low coupling values.
It is also possible for DAMSA to explore the $\mu^{+}\mu^{-}$ final state of the ALPs, given the 400GeV beam energy, which will benefit from an addition of a muon detection system.

We anticipate that DAMSA could possibly probe the region of ALP parameter space comparable or slightly broader in the upper-right corner of the phase space, clearly demonstrating SHiP's physics case. In this regard, DAMSA can be regarded as a complementary example case to SHiP, demonstrating the possibility of exploring parameter space regions in a relatively short exposure time.  
As SHiP extends its reach into the lower-right corner of the phase space by collecting more data, DAMSA and SHiP will provide complementary information in the search for ALPs.

\section{The Next Steps, The Little DAMSA Path Finder}

The use of high-intensity proton beams and the close proximity of the detector inevitably lead to a high flux of low-energy neutrons, which may induce unwanted backgrounds.  DAMSA aims to overcome this challenge by selecting specific final states, optimizing beam configurations, and employing a detector system designed to mitigate neutron backgrounds. 
To demonstrate the physics case of DAMSA, we plan on constructing and operating the {\bf Little DAMSA Path-Finder (LDPF)} proof-of-concept experiment within the next 1--2 years, namely by mid 2027. This timeline provides ample time flexibility to prepare a full scale experiment with any necessary modifications within this decade.
LDPF will focus on the two-photon final state of the ALPs and will take data, using the 300~MeV electron beams---potentially offering a more controlled environment regarding neutron-induced backgrounds---at Fermilab’s Facility for Accelerator Science and Technology (FAST) and the detector system placed $\sim10$~cm away from the ALP production point in the target. 

A successful completion of the LDPF will enable us not only to clearly demonstrate the accessibility of the targeted parameter space but also to build confidence in the experimental techniques, paving the way for the full-scale DAMSA experiment at the proton facilities, in ample time for data taking even during the CERN BDF and SHiP experiment commissioning in early 2030's.

\section{Conclusions}

Our detailed studies clearly show that DAMSA's small scale and the extreme short baseline nature enables accessing the parameter space that has been challenging to access for the benchmark physics case of ALPs.
In particular, regarding DSP mediator searches (e.g., ALPs, dark photons), especially for the relatively short-lived ones, the SHiP experiment operating with the BDF facility at CERN is expected to offer unique opportunities for exploring parameter space regions with relatively large couplings and mass values. This advantage stems from SHiP’s high beam energy and close proximity to the signal production point (i.e., the beam target), compared to other experimental setups. 
In this white paper, we propose to position the DAMSA detector immediate downstream of the muon shield, ensuring and complementing the experiment's sensitivity to relatively short-lived mediators without interfering with the operation of other experiments at BDF.
Given the cost effective, table-top scale dimension of the experiment and the timeline presented above, it is entirely possible for DAMSA to be ready to take data even during the CERN BDF and SHiP experiment commissioning in early 2030's.

\section*{Acknowledgments}
The work of JY is partially supported by the University of Texas at Arlington, U.S. Department of Energy under Grant No. DE-SC0011686 and the National Research Foundation grant NRF-RS-2024-00350406.
The work of DK is supported partially by the University of South Dakota.
The work of UKY is partially supported by the National Research Foundation grant NRF-RS-2024-00350406.

\bibliography{euroStrat}

\end{document}